\documentclass[12pt]{article}
\usepackage{amssymb,amsfonts}
\usepackage{graphics,psboxit,amsmath}
\usepackage{subfigure}
\usepackage{graphicx}
\usepackage{verbatim}
\usepackage{color}
\usepackage{hyperref}
\hypersetup{colorlinks}

\definecolor{darkred}{rgb}{1,0,0}
\definecolor{darkgreen}{rgb}{0,0.5,0}
\definecolor{darkblue}{rgb}{0,0,1}
\definecolor{orange}{rgb}{1,0.5,0}
\definecolor{green}{rgb}{0,1,0}
\definecolor{purple}{rgb}{.5,0,1}

\hypersetup{ colorlinks,
linkcolor=darkred,
filecolor=darkgreen,
urlcolor=darkblue,
citecolor=darkblue,
linktocpage=true }

\definecolor{markcolor}{rgb}{.25,0,1}

\definecolor{markcolor2}{rgb}{1,0,0}

\definecolor{markcolor3}{rgb}{0,1,0}

\def\hybrid{\topmargin 0pt    \oddsidemargin 0.05in 
        \headheight 0pt \headsep 0pt
        \textwidth 16.3cm      
        \textheight 22,5cm       
        \marginparwidth .875in
        \parskip 5pt plus 1pt   \jot = 1.5ex}

\hybrid

\catcode`\@=11

\def\marginnote#1{}
\newcount\hour
\newcount\minute
\newtoks\amorpm
\hour=\time\divide\hour by60 \minute=\time{\multiply\hour by60
\global\advance\minute by-\hour}
\edef\standardtime{{\ifnum\hour<12 \global\amorpm={am}%
        \else\global\amorpm={pm}\advance\hour by-12 \fi
        \ifnum\hour=0 \hour=12 \fi
        \number\hour:\ifnum\minute<10 0\fi\number\minute\the\amorpm}}
\edef\militarytime{\number\hour:\ifnum\minute<10 0\fi\number\minute}

\def\draftlabel#1{{\@bsphack\if@filesw {\let\thepage\relax
   \xdef\@gtempa{\write\@auxout{\string
      \newlabel{#1}{{\@currentlabel}{\thepage}}}}}\@gtempa
   \if@nobreak \ifvmode\nobreak\fi\fi\fi\@esphack}
        \gdef\@eqnlabel{#1}}
\def\@eqnlabel{}
\def\@vacuum{}
\def\draftmarginnote#1{\marginpar{\raggedright\scriptsize\tt#1}}

\def\draft{\oddsidemargin -.5truein
        \def\@oddfoot{\sl preliminary draft \hfil
        \rm\thepage\hfil\sl\today\quad\militarytime}
        \let\@evenfoot\@oddfoot \overfullrule 3pt
        \let\label=\draftlabel
        \let\marginnote=\draftmarginnote
   \def\@eqnnum{(\theequation)\rlap{\kern\marginparsep\tt\@eqnlabel}%
\global\let\@eqnlabel\@vacuum}  }

\def\draft2{
        \def\@oddfoot{\sl preliminary draft \hfil
        \rm\thepage\hfil\sl\today\quad\militarytime}
        \let\@evenfoot\@oddfoot \overfullrule 3pt
        \let\label=\draftlabel
        \let\marginnote=\draftmarginnote
   \def\@eqnnum{(\theequation)\rlap{\kern\marginparsep\tt\@eqnlabel}%
\global\let\@eqnlabel\@vacuum}  }

\def\preprint{\twocolumn\sloppy\flushbottom\parindent 2em
        \leftmargini 2em\leftmarginv .5em\leftmarginvi .5em
        \oddsidemargin -.5in    \evensidemargin -.5in
        \columnsep .4in \footheight 0pt
        \textwidth 10.in        \topmargin  -.4in
        \headheight 12pt \topskip .4in
        \textheight 6.9in \footskip 0pt
        \def\@oddhead{\thepage\hfil\addtocounter{page}{1}\thepage}
        \let\@evenhead\@oddhead \def\@oddfoot{} \def\@evenfoot{} }

\def\numberbysection{\@addtoreset{equation}{section}
        \def\theequation{\thesection.\arabic{equation}}}

\def\underline#1{\relax\ifmmode\@@underline#1\else
        $\@@underline{\hbox{#1}}$\relax\fi}

\def\titlepage{\@restonecolfalse\if@twocolumn\@restonecoltrue\onecolumn
     \else \newpage \fi \thispagestyle{empty}\c@page\z@
        \def\thefootnote{\fnsymbol{footnote}} }

\def\endtitlepage{\if@restonecol\twocolumn \else \newpage \fi
        \def\thefootnote{\arabic{footnote}}
        \setcounter{footnote}{0}}  

\catcode`@=12 \relax

\def\figcap{\section*{Figure Captions\markboth
        {FIGURECAPTIONS}{FIGURECAPTIONS}}\list
        {Figure \arabic{enumi}:\hfill}{\settowidth\labelwidth{Figure
999:}
        \leftmargin\labelwidth
        \advance\leftmargin\labelsep\usecounter{enumi}}}
 \relax
\def\tablecap{\section*{Table Captions\markboth
        {TABLECAPTIONS}{TABLECAPTIONS}}\list
        {Table \arabic{enumi}:\hfill}{\settowidth\labelwidth{Table
999:}
        \leftmargin\labelwidth
        \advance\leftmargin\labelsep\usecounter{enumi}}}
 \relax
\def\reflist{\section*{References\markboth
        {REFLIST}{REFLIST}}\list
        {[\arabic{enumi}]\hfill}{\settowidth\labelwidth{[999]}
        \leftmargin\labelwidth
        \advance\leftmargin\labelsep\usecounter{enumi}}}
 \relax
\makeatletter
\newcounter{pubctr}
\def\publist{\@ifnextchar[{\@publist}{\@@publist}}
\def\@publist[#1]{\list
        {[\arabic{pubctr}]\hfill}{\settowidth\labelwidth{[999]}
        \leftmargin\labelwidth
        \advance\leftmargin\labelsep
        \@nmbrlisttrue\def\@listctr{pubctr}
        \setcounter{pubctr}{#1}\addtocounter{pubctr}{-1}}}
\def\@@publist{\list
        {[\arabic{pubctr}]\hfill}{\settowidth\labelwidth{[999]}
        \leftmargin\labelwidth
        \advance\leftmargin\labelsep
        \@nmbrlisttrue\def\@listctr{pubctr}}}
 \relax
\makeatother

\def\be{\begin{equation}}
\def\ee{\end{equation}}
\def\ba{\begin{eqnarray}}
\def\ea{\end{eqnarray}}

\def\no{\noindent}

\def\IR{\relax{\rm I\kern-.18em R}}

\def\bse{\begin{small}\begin{equation*}}
\def\ese{\end{equation*}\end{small}}

\begin{document}


\renewcommand{\theequation}{\thesection.\arabic{equation}}
\csname @addtoreset\endcsname{equation}{section}

\newcommand{\eqn}[1]{(\ref{#1})}

\begin{titlepage}
\begin{center}
\strut\hfill
\vskip 1.3cm


\vskip .5in

{\Large \bf Lax pair formulation in the simultaneous presence of boundaries and defects}

\vskip 0.5in

{\large \bf Anastasia Doikou}\phantom{x}
\vskip 0.02in
{\footnotesize Department of Mathematics, Heriot-Watt University,\\
EH14 4AS, Edinburgh, United Kingdom}
\\[2mm]
\noindent
{\footnotesize and}

\vskip 0.02in
{\footnotesize Department of Computer Engineering \& Informatics,\\
 University of Patras, GR-Patras 26500, Greece}
\\[2mm]
\noindent
\vskip .1cm


{\footnotesize {\tt E-mail: A.Doikou@hw.ac.uk}}\\

\end{center}

\vskip 1.0in

\centerline{\bf Abstract}

Inspired by recent results on the effect of integrable boundary conditions on the bulk behavior of an integrable system, and in particular on the behavior of an existing defect we systematically formulate the Lax pairs in the simultaneous presence of integrable boundaries and defects. The respective sewing conditions as well as the relevant equations of motion on the defect point are accordingly extracted. We consider a specific prototype i.e. the vector non-linear Schr\"{o}dinger (NLS) model to exemplify our construction. This model displays a highly non-trivial behavior and allows the existence of two distinct types of boundary conditions based on the reflection algebra or the twisted Yangian.
\no

\vfill
\end{titlepage}
\vfill \eject

\tableofcontents

\section{Introduction}

The main objective of the present investigation is the construction of Lax pairs in the simultaneous presence of integrable boundaries and defects. This study is mainly inspired by recent results on the study of integrable impurities at both clasical and quantum level \cite{delmusi}--\cite{doikou-defects} as well as the findings of \cite{doikou-vectornls}, where the presence of non trivial boundary conditions yielded highly non-trivial behavior associated to the defect.
We shall provide here the generic framework for the defect in the presence of two distinct types of boundary conditions, and shall exemplify our construction using the vector non-linear Schr\"{o}dinger (NLS) model. The underlying high rank algebra $\mathfrak{gl}_N$ that rules the model allows the existence of a rich variety of boundary conditions that preserve integrability, which renders the behavior of the system even more intricate. It is worth noting that in addition to the derivation of the local integrals of motion presented in \cite{doikou-vectornls} the Lax pair construction is equally important providing the sewing conditions compatible with integrability. They are entailed as analyticity requirements imposed on the time components of the Lax pair around the defect point (see also e.g. \cite{avan-doikou1}).
The derivation of Lax pairs, which is one of the main results in this context --in the simultaneous presence of bounaries and defcts-- and is presented in section 2, is the starting point for making contact with the B\"{a}cklund transformation on the defect point (see e.g. \cite{BZ}). The present investigation may be seen as the continuation of the work in \cite{doikou-vectornls}, where the local integrals of motion were extracted. Here explicit derivation and computation of Lax pairs, derivation of the sewing conditions and the equations of motion associated to the defect point is presented.

Let us introduce below the general setting of our formulation. We shall consider here the situation where both non-trivial boundary conditions and point-like defects are present in the 1+1 dimensional classical field theory. Let us first introduce the modified monodromy matrix to include the point-like defect at $x=x_0$ (see e.g. \cite{ft, BCZ1, avan-doikou1} and references therein):
\be
T(\lambda; L, 0) = T^{+}(L, x_0; \lambda)\ {\mathbb L}(x_0; \lambda)\ T^{-}(x_0, 0; \lambda) \label{modT}
\ee
where ${\mathbb L}$ is the algebraic object associated to the local defect and is required to be a representation of a suitable quadratic classical algebra as will be discussed later in the text, we also define the ``bulk'' monodromy matrices as (see e.g. \cite{ft}):
\be
T^{\pm}(x, y; \lambda) = {\cal P}\exp\{ \int_{y}^x dx'\  {\mathbb U}^{\pm}(x'; \lambda)\} \label{tpm}
\ee
and they are solutions of the differential equation
\ba
{\partial T^{\pm}(x,y)\over
\partial x} = {\mathbb U}^{\pm}(x,t, \lambda) T^{\pm}(x,y). \label{dif0}
\ea
The fact that $T^{\pm}$ is
a solution of equation (\ref{dif0}) will be extensively used
subsequently for obtaining the relevant integrals of motion.
${\mathbb U}^{\pm}$ is part of the Lax pair ${\mathbb U}^{\pm},\ {\mathbb V}^{\pm}$ that satisfy the linear auxiliary problem.
Let $\Psi$ be a solution of the following set of
equations
\ba
&&{\partial \Psi \over
\partial x} = {\mathbb U}^{\pm}(x,t, \lambda) \Psi \label{dif1}
\\ &&
{\partial  \Psi \over \partial t } = {\mathbb V}^{\pm}(x,t,\lambda) \Psi
\label{dif2}
\ea
${\mathbb U}^{\pm},\ {\mathbb V}^{\pm}$ are in general
$n \times n$ matrices with entries functions of complex valued
fields, their derivatives, and the spectral parameter $\lambda$.
Compatibility conditions of the two differential equations
(\ref{dif1}), (\ref{dif2}) lead to the zero curvature condition
\be
\dot{{\mathbb U}^{\pm}} - {\mathbb V}^{\pm'} + \Big [{\mathbb U}^{\pm},\ {\mathbb V}^{\pm}
\Big ]=0, \label{zecu}
\ee
giving rise to the corresponding
classical equations of motion of the system under consideration. Special care should be taken on the defect point; the zero curvature condition on the point reads as \cite{avan-doikou1, avan-doikou}
\be
{d {\mathbb L}(\lambda) \over d t} = \tilde {\mathbb V}^+(\lambda)\ {\mathbb L}(\lambda) - {\mathbb L}(\lambda)\ \tilde {\mathbb V}^-(\lambda) \label{zcd}
\ee
where $\tilde {\mathbb V}^{\pm}$ are the time components of the Lax pairs on the defect point form the left (right). This is an intricate equation, which arises naturally when studying the continuum limit of discrete integrable theories (see e.g. \cite{avan-doikou1}).

Let us also briefly review the algebraic setting regarding the system in the presence of point like defects and integrable boundaries, i.e. the Hamiltonian formalism.
The existence of the classical $r$-matrix \cite{sts}, satisfying the classical Yang-Baxter equation
\be
\Big [r_{12}(\lambda_1-\lambda_2),\
r_{13}(\lambda_1)+r_{23}(\lambda_2) \Big]+ \Big
[r_{13}(\lambda_1),\ r_{23}(\lambda_2) \Big] =0,
\ee guarantees
the integrability of the classical system.
The monodromy matrices $T^{\pm}$ as well as the defect matrix ${\mathbb L}$ and consequently the modified monodromy matrix $T(L,0;\lambda)$ satisfy the classical quadratic algebra \cite{ft}:
\be \Big
\{T^{\pm}_{1}(x,y,t;\lambda_1),\ T^{\pm}_{2}(x,y,t;\lambda_2) \Big \} =
\Big[r_{12}(\lambda_1-\lambda_2),\
T^{\pm}_1(x,y,t;\lambda_1)\ T^{\pm}_2(x,y,t;\lambda_2) \Big ]. \label{basic}
\ee
also the main requirement \cite{avan-doikou1} for the ${\mathbb L}$-matrix in (\ref{modT}) so that integrability is ensured is:
\be \Big
\{{\mathbb L}_{1}(\lambda_1),\ {\mathbb L}_{2}(\lambda_2) \Big \} =
\Big[r_{12}(\lambda_1-\lambda_2),\
{\mathbb L}_1(\lambda_1)\ {\mathbb L}_2(\lambda_2) \Big ]. \label{rtt2}
\ee
and
\be
\{T_1^+(\lambda_1),\ T_2^-(\lambda_2) \} = 0.
\ee

Here, we are interested as already mentioned in the situation when both local defects and integrable boundary conditions are present. As in \cite{doikou-vectornls} we shall distinguish two cases of boundary conditions the
soliton preserving (SP) associated to the reflection algebra and the soliton non-preserving (SNP) associated to the twisted Yangian:
\\
\\
$\bullet$ {\bf Reflection algebra}: The classical reflection algebra \cite{sklyanin} describes the so called soliton-preserving (SP) boundary conditions:
\ba
\Big \{ {\cal T}_1(\lambda),\  {\cal T}_2(\mu) \Big \} &=& r_{12}(\lambda-\mu)\ {\cal T}_1(\lambda)\ {\cal T}_2(\mu) - {\cal T}_1(\lambda)\ {\cal T}_2(\mu)\ r_{21}(\lambda-\mu) \cr &+& {\cal T}_1(\lambda)\ r_{21}(\lambda+\mu)\ {\cal T}_2(\mu)\ - {\cal T}_2(\mu)\ r_{12}(\lambda+\mu)\ {\cal T}_1(\lambda) \label{ra}
\ea
The corresponding representation of the algebra is given as \cite{sklyanin}:
\be
{\cal T}_{a}(\lambda) = T_a(L, 0; \lambda)\ K_a(\lambda)\ T_a^{-1}(L, 0; -\lambda) \label{rep1}
\ee
where the monodromy matrix includes also the defects and is expressed as in (\ref{modT}).
Recall  that $T^{\pm}$ are defined in (\ref{tpm}). Also $K$ is a $c$-number representation
\be
\{K_a(\lambda),\ K_b(\mu)\} =0. \label{nondyn}
\ee
of the classical reflection algebra (\ref{ra}).
The local integrals of motion are obtained via the generating function
\be
{\cal G}(\lambda) = \ln tr_a(\bar K_a(\lambda){\cal T}_a(\lambda)), \label{gener}
\ee
$\bar K$ is also a c-number (non-dynamical) (\ref{nondyn}) solution of the classical reflection equation
\\
\\
$\bullet$ {\bf Twisted Yangian}:
The classical twisted Yangian \cite{sklyanin, olshanski} describes the soliton non-preserving boundary conditions (SNP) and is defined as:
\ba
\Big \{ {\cal T}_1(\lambda),\  {\cal T}_2(\mu) \Big \} &=& r_{12}(\lambda-\mu)\ {\cal T}_1(\lambda)\ {\cal T}_2(\mu) - {\cal T}_1(\lambda)\ {\cal T}_2(\mu)\ r_{21}(\lambda-\mu) \cr &+& {\cal T}_1(\lambda)\ r_{21}^{t_{1}}(\lambda+\mu)\ {\cal T}_2(\mu)\ - {\cal T}_2(\mu)\ r_{12}^{t_2}(\lambda+\mu)\ {\cal T}_1(\lambda) \label{ty}
\ea
$^{t_i}$ denotes transposition on the auxiliary space $i$.
The generic representation of the classical twisted Yangian is given by \cite{sklyanin, doikou-twisted}
\be
{\cal T}_a(\lambda) =  T_a(L ,0; \lambda)\ K_a(\lambda)\ T_a^{t_a}(L, 0; -\lambda), \label{rep2}
\ee
where $K$ is a non-dynamical solution of the reflection equation (\ref{nondyn}).
The local integrals of motion are obtained via the generating function (\ref{gener}).
Having set the algebraic background we may now proceed with the construction of the Lax pairs.

\section{The Lax pair construction}

We shall focus on the SP case mainly because in this case non-trivial bulk behavior is manifest.
Nevertheless, we shall briefly describe the SNP case as well in the subsequent section. Note that the basic expressions derived here are to our knowledge novel. We shall first focus on the construction of the time component of the Lax pairs for the left and right bulk theories. To derive the time component of the Lax pair in the simultaneous presence of boundaries and defects we shall need the following fundamental algebraic relations (see also \cite{ft,avan-doikou1, avan-doikou-lax}, for more details):
\be
\Big \{ T_a(\lambda),\ {\mathbb U}_b^{\pm}(\mu)\Big \} = {\partial M_{ab}^{\pm}(\lambda, \mu) \over \partial x} + \Big [M_{ab}^{\pm}(\lambda, \mu),\ {\mathbb U}^{\pm}_b(\mu) \Big ], \label{al1}
\ee
where
\ba
&& M_{ab}^+(\lambda, \mu) = T_a^+(L, x;\lambda)\ r_{ab}(\lambda-\mu)\ T_a^+(x, x_0; \lambda)\ {\mathbb L}_a(x_0; \lambda)\ T_a^-(x_0, 0; \lambda), ~~~~x > x_0 \cr
&& M_{ab}^-(\lambda, \mu) =  T_a^+(L, x_0; \lambda)\ {\mathbb L}_a(x_0; \lambda)\ T_a^-(x_0, x; \lambda)\ r_{ab}(\lambda-\mu)\ T_a^-(x_0, 0; \lambda) ~~~~x < x_0.  \nonumber\\ \label{m1}
\ea
Recall that here we are interested  in the SP (reflection algebra) boundary conditions.
In order to formulate the suitable Poisson bracket for our purposes we shall also need the following
\be
\Big \{ T^{-1}_a(-\lambda),\ {\mathbb U}_b^{\pm}(\mu)\Big \} = {\partial M^{\pm *}_{ab}(\lambda, \mu) \over \partial x} + \Big [M_{ab}^{\pm *}(\lambda, \mu),\ {\mathbb U}^{\pm}_b(\mu) \Big ] \label{al2}
\ee
where we define:
\ba
&& M_{ab}^{+*}(\lambda, \mu)=  T_a^-(0, x;-\lambda)^{-1}\ {\mathbb L}_a^{-1}(x_0;-\lambda)\ r_{ab}(\lambda+\mu)\  T_a^+(x, x_0; -\lambda)^{-1}\ T_a^+(L, x; -\lambda)^{-1}, ~~~~x > x_0\cr
&& M_{ab}^{-*}(\lambda, \mu) =  T_a^-(x, 0; -\lambda)^{-1}\ r_{ab}(\lambda+\mu )\ T_a^-(x_0, x; -\lambda)^{-1}\ {\mathbb L}_a^{-1}(x_0; -\lambda)\ T_a^+(L, x_0; -\lambda)^{-1}, ~~~~x < x_0. \nonumber\\ \label{m*1}
\ea
Using the fundamental relations (\ref{al1}), (\ref{al2}) we can formulate the necessary for our purposes Poisson bracket
\be
\Big \{\bar K_a(\lambda){\cal T}_a(\lambda),\ {\mathbb U}^{\pm}_b(\mu) \Big \} = {\partial \over \partial x}\Big ({\cal M}^{\pm}_{ab}(\lambda,\mu) + {\cal M}_{ab}^{\pm *}(\lambda,\mu) \Big ) + \Big [ {\cal M}^{\pm}_{ab}(\lambda, \mu)+ {\cal M}_{ab}^{\pm*}(\lambda, \mu) ,\ {\mathbb U}_{b}(\mu)\Big ]
\ee
and ${\cal M}^{\pm},\ {\cal M}^{\pm *}$ are defined as:
\ba
&& {\cal M}_{ab}^{\pm}(\lambda, \mu) = \bar K_a(\lambda)\ M_{ab}^{\pm}(\lambda, \mu)\ K_a(\lambda)\ \ T_a^{-1}(-\lambda) \cr
&& {\cal M}_{ab}^{\pm *}(\lambda, \mu) = \bar K_a(\lambda)\ T_a(\lambda)\  K_a(\lambda)\ M_{ab}^{\pm*}(\lambda,\mu).
\ea
$M^{\pm},\ M^{\pm *}$ are defined in (\ref{m1}), (\ref{m*1}).

After considering the generating function of the local integrals of motion ${\cal G}(\lambda)$ we conclude that (see also \cite{ft}):
\be
\dot {\mathbb U}^{\pm}(\mu) = {\partial {\mathbb V}^{\pm}(\lambda, \mu) \over \partial x } +
\Big [ {\mathbb V}^{\pm}(\lambda, \mu),\ {\mathbb U}^{\pm}(\mu) \Big ]
\ee
where the bulk quantities ${\mathbb V}^{\pm}$ are expressed as (they are defined up to an overall normalization factor):
\be
{\mathbb V}_b^{\pm}(\lambda, \mu; x) \propto t^{-1}(\lambda)\ tr_a \Big ({\cal M}_{ab}^{\pm}(\lambda, \mu) +
{\cal M}_{ab}^{\pm *}(\lambda, \mu) \Big ) ~~~~~x \neq x_0 \label{VV}
\ee
where we define
\be
t(\lambda) = tr_a (\bar K_a(\lambda)\ {\cal T}_a(\lambda)) \label{t0}
\ee
${\cal T}$ is defined in (\ref{rep1}) in the reflection algebra case, and in (\ref{rep2}) in the twisted Yangian case. Expansion of the ${\mathbb V}^{\pm}$ in powers of $\lambda^{-1}$ will provide the time components of the Lax pairs associated to each local integral of motion
\be
{\mathbb V}^{\pm}(\lambda, \mu) = \sum_n {{\mathbb V}^{\pm(n)}(\mu)\over \lambda^n}. \label{expand1}
\ee

As already mentioned earlier in the text special care is taken on the defect point. Taking into account the ${\mathbb L}$-matrix satisfies the quadratic algebra (\ref{rtt2}) we conclude:
\be
\Big \{T_a(\lambda),\ {\mathbb L}_b(\mu) \Big \} =
\tilde M_{ab}^+(\lambda, \mu)\ {\mathbb L}_b(\mu) - {\mathbb L}_b(\mu)\ \tilde M_{ab}^-(\lambda, \mu)
\ee
and
\be
\Big \{ T_a^{-1}(-\lambda),\ {\mathbb L}_b(\mu)\Big \} =
\tilde M_{ab}^{+*}(\lambda, \mu)\ {\mathbb L}_b(\mu) - {\mathbb L}_b(\mu)\ \tilde M_{ab}^{-*}(\lambda, \mu).
\ee
where we define
\ba
&& \tilde M^+_{ab}(\lambda, \mu) = T^+_{a}(L,x_0; \lambda)\ r_{ab}(\lambda - \mu)\ {\mathbb L}_a(\lambda)\ T^-_a(x_0, 0 ; \lambda) \cr
&& \tilde  M^-_{ab}(\lambda, \mu) = T^+_{a}(L, x_0; \lambda)\ {\mathbb L}_a(\lambda)\ r_{ab}(\lambda - \mu)\ T^-_a(x_0, 0 ; \lambda) \cr
&& \tilde M^{+*}_{ab}(\lambda, \mu) =  T_a^-(x_0, 0; -\lambda)^{-1}\  {\mathbb L}_a^{-1}(-\lambda)\ r_{ab}(\lambda+\mu)\  T_a^+(L,0; -\lambda)^{-1} \cr
&& \tilde M^{-*}_{ab}(\lambda, \mu) =  T_a^-(x_0, 0;-\lambda)^{-1}\  r_{ab}(\lambda+\mu)\
{\mathbb L}^{-1}_a(-\lambda)\  T_a^+(L,0; - \lambda)^{-1}.
\ea
Then it is straightforward to show that
\be
\Big \{\bar K_a(\lambda)\ {\cal T}_a(\lambda),\ {\mathbb L}_b(\mu) \Big \} =\Big ( \tilde {\cal M}^{+ }_{ab}(\lambda,\mu)+ \tilde {\cal M}_{ab}^{+*}(\lambda, \mu)\Big )\ {\mathbb L}_b(\mu) - {\mathbb L}_b(\mu)
\ \Big (\tilde {\cal M}^{-}_{ab}(\lambda, \mu) + \tilde {\cal M}_{ab}^{-*}(\lambda, \mu)\Big)
\ee
where
\ba
&& \tilde {\cal M}_{ab}^{\pm}(\lambda, \mu) = \bar K_a(\lambda)\ \tilde M^{\pm}_{ab}(\lambda, \mu)\ K_a(\lambda)\ T_a^{-1}(-\lambda)  \cr
&&  \tilde {\cal M}_{ab}^{\pm*}(\lambda, \mu) = \bar K_a(\lambda)\ T_a(\lambda)\  K_a(\lambda)\ \tilde M^{\pm *}_{ab}(\lambda, \mu).
\ea

The aim now is to identify $\tilde {\mathbb V}^{\pm}(x_0)$, i.e. the Lax pairs on the defect point from the left and from the right. Indeed, bearing in mind the zero curvature condition on the defect point we may directly identify the relevant $\tilde {\mathbb V}^{\pm}$ operators (defined up to an overall normalization factor)
\be
\tilde {\mathbb V}_b^{\pm}(\lambda, \mu; x_0) \propto t^{-1}(\lambda)\ tr_a \Big ( \tilde {\cal M}^{\pm }_{ab}(\lambda, \mu) + \tilde {\cal M}^{\pm *}_{ab}(\lambda, \mu)\Big ). \label{tvv}
\ee
As in the bulk case $\tilde {\mathbb V}^{\pm}(\lambda,\mu) = \sum_{n} {\tilde {\mathbb V}^{\pm(n)}(\mu) \over \lambda^n}$.

Then analyticity conditions imposed on the time components of the Lax pairs
\be
\tilde {\mathbb V}^{\pm(n)}(x_0^{\pm}) \to {\mathbb V}^{\pm(n)}(x_0) \label{cont}
\ee
will provide the wanted sewing conditions as explained in \cite{avan-doikou1}.

Note that in the case of SNP boundary conditions the same expressions for the time components are valid, but now we define
\ba
&& M^{\pm *}_{ab}(\lambda, \mu) = \Big ( M_{ab}^{\pm}(-\lambda,\mu)\Big )^{t_a} \cr
&& \tilde M^{\pm*}_{ab}(\lambda, \mu) = \Big ( \tilde M_{ab}^{\pm}(-\lambda,\mu)\Big )^{t_a}.
\ea

In the next section we shall consider the vector NLS model in the presence of defects and boundaries associated to the reflection equation and the twisted Yangian.

\section{The vector NLS theory}

We shall focus hereafter on the $\mathfrak{gl}_N$ vector NLS model.
Let us first review the necessary information regarding the model.
The Lax pair for the generalized NLS model is given by the following expressions
\cite{ft}:
\be
{\mathbb U} = {\mathbb U}_0 + \lambda
{\mathbb U}_1, ~~~{\mathbb V} = {\mathbb V}_0+\lambda{\mathbb V}_1
+\lambda^2 {\mathbb V}_2, \label{UV1}
\ee
where
\ba
&&
{\mathbb U}_1 = {1\over 2i} (\sum_{i=1}^{N-1}e_{ii} -e_{NN}),
~~~~{\mathbb U}_0 =
\sqrt{\kappa} \sum_{i=1}^{N-1}(\bar \psi_i e_{iN} +\psi_i e_{Ni}) \cr
&& {\mathbb V}_0 = i\kappa \sum_{i,\ j=1}^{N-1}(\bar \psi_i \psi_j
e_{ij} -|\psi_i|^2 e_{NN}) -i\sqrt{\kappa}\sum_{i=1}^{N-1} (\bar
\psi_i' e_{iN} - \psi_i' e_{Ni}), \cr
&&
{\mathbb V}_1= -{\mathbb U}_{0}, ~~~{\mathbb V}_2= -{\mathbb U}_{1}. \label{lax}
\ea
$e_{ij}$ are $ N \times N$ matrices such that: $~(e_{ij})_{kl} = \delta_{ik} \delta_{jl}$.
Consider also the corresponding classical $r$-matrix \cite{sklyanin2, yang}
\be
r(\lambda) = {\kappa {\mathrm P} \over \lambda}
~~~~\mbox{where} ~~~~{\mathrm P}=\sum_{i,j=1}^{N} e_{ij} \otimes
e_{ji} \label{rr}
\ee
${\mathrm P}$ is the $\mathfrak{gl}_N$ permutation operator.

The fields $\psi_i,\ \bar \psi_j$ satisfy\footnote{The Poisson
structure for the generalized NLS model is defined as:
\be
\Big \{
A,\ B  \Big \}=  -i \sum_{i} \int_{-L}^{L} dx \Big ({\delta A \over
\delta \psi_i(x)}\ {\delta B \over \delta \bar \psi_i(x)} -
{\delta A \over \delta \bar \psi_i(x)}\ {\delta B \over \delta
\psi_i(x)}\Big )
\ee}:
\be
\Big \{ \psi_{i}(x),\ \psi_j(y) \Big
\} = \Big \{\bar \psi_{i}(x),\ \bar \psi_j(y) \Big \} =0, ~~~~\Big
\{\psi_{i}(x),\ \bar \psi_j(y) \Big \}= -i \delta_{ij}\ \delta(x-y).
\ee
From the zero curvature condition (\ref{zecu}) the familiar classical
equations of motion for the generalized NLS model are entailed \cite{ft}.

To obtain the local integrals of motion of the
NLS model one has to expand the monodromy matrix in powers of
$\lambda^{-1}$ \cite{ft}. Let us consider the following ansatz for
$T$ as $|\lambda| \to \infty$
\be
T(x,y,\lambda) = ({\mathbb I}
+W(x, \lambda))\ \exp[Z(x,y,\lambda)]\ ({\mathbb I}
+W(y,\lambda))^{-1} \label{exp0}
\ee
where $W$ is off diagonal
matrix i.e. $~W = \sum_{i\neq j} W_{ij} e_{ij}$, and $Z$ is purely
diagonal $~Z = \sum_{i=1}^N Z_{ii}e_{ii}$.
Also
\be
Z_{ii}(\lambda) = \sum_{n=-1}^{\infty} {Z^{(n)}_{ii} \over
\lambda^{n}}, ~~~~W_{ij} = \sum_{n=1}^{\infty}{W_{ij}^{(n)} \over
\lambda^n}. \label{expa}
\ee
Inserting the latter expressions
(\ref{expa}) in (\ref{dif1}) one may identify the coefficients
$W_{ij}^{(n)}$ and $Z_{ii}^{(n)}$. Notice that as $i\lambda \to \infty$ the only non
negligible contribution from $Z^{(n)}$ comes from the
$Z^{(n)}_{NN}$ term, and is given by:
\be Z_{NN}^{(n)}(L, -L)= i L
\delta_{n, -1} +\sqrt{\kappa}\sum_{i=1}^{N-1}\int_{-L}^L dx\
\psi_i(x)\ W_{iN}^{(n)}(x). \label{ref1}
\ee
It is thus sufficient
to determine the coefficients $W_{iN}^{(n)}$ in order to extract
the relevant local integrals of motion.
Indeed solving (\ref{dif1}) one may easily obtain (see also \cite{ft} for more details on the relevant computations):
\ba &&
W_{iN}^{(1)}(x) = -i \sqrt{\kappa} \bar \psi_{i}(x),
~~~~W_{iN}^{(2)}(x)=\sqrt{\kappa} \bar \psi_i'(x), ~~~~~\mbox{also} \label{ref2}\cr
&&
W_{Ni}^{(1)}(x) = i\sqrt{\kappa}\psi_i(x), ~~~~W_{Ni}^{(2)}(x) = -i
W^{'(1)}_{Ni}(x) +\sum_{i\neq j,\ i, j \neq N}W^{(1)}_{N
j}(x)W_{ji}^{(1)}(x).
\label{reff}
\ea

Note that we choose to consider as a defect matrix the following:
\be
{\mathbb L}(\lambda; x_0) = \lambda + \kappa {\mathbb P}(x_0), ~~~~~
{\mathbb P} = \sum_{k,l =1}^{N} {\mathbb P}_{kl}\ e_{kl} \label{ll1}
\ee
where ${\mathbb P}_{kl}$ satisfy the ultra-local $
\mathfrak{gl}_N$ classical algebra \cite{ft}:
\be
\Big \{ {\mathbb P}_{ij}(x_0),\ {\mathbb P}_{kl}(x_0) \Big \} = {\mathbb P}_{il}(x_0) \delta_{jm} - {\mathbb P}_{kj}(x_0)\delta_{il}. \label{ll2}
\ee

\subsection{The reflection algebra}

To identify the time components of the Lax pairs associated to each local integral of motion we expand the expressions for ${\mathbb V}^{\pm},\ \tilde {\mathbb V}^{\pm}$ in powers of ${1 \over \lambda}$:
\be
{\mathbb V}^{\pm}(\lambda, \mu) = \sum_{n=1}^{\infty} { {\mathbb V}^{\pm(n)}(\mu) \over \lambda^n},
~~~~\tilde {\mathbb V}^{\pm}(\lambda, \mu) = \sum_{n=1}^{\infty} {\tilde {\mathbb V}^{\pm(n)}(\mu) \over \lambda^n}.
\ee
Recall, that in the SP case only the odd integrals of motions are the conserved charges (see e.g. \cite{sklyanin}) . Computing the bulk ${\mathbb V}^{\pm}$ operators, associated to the first two non-trivial integrals of motion i.e. number of particles and Hamiltonian, for the right and left theories we recover the familiar bulk components:
\be
{\mathbb V}^{\pm(1)} = e_{NN},
\ee
whereas ${\mathbb V}^{\pm(3)}$ --relevant to the Hamiltonian-- are given in (\ref{UV1}), (\ref{lax}) ($\psi,\ \bar \psi\ \to\ \psi^{\pm},\ \bar \psi^{\pm}$), and $\tilde {\mathbb V}^{(2)}$, associated to the momentum is zero, given that the momentum is not a conserved quantity anymore.

To extract the desired sewing condition in the SP case we focus on the computation of the operators on the defect point and impose the suitable analyticity condition from the left and the right. Moreover, having this information at our disposal we will be able to obtain the associated equations of motion on the defect point.
The $\tilde {\mathbb V}^{\pm}$ operators associated to the first three orders of the expansion are given below:
\ba
&&\tilde {\mathbb V}^{\pm (1)}(\mu) = e_{NN} \cr
&& \tilde {\mathbb V}^{\pm (2)}(\mu) = {1\over 2} ( {\cal A}^{\pm} + \bar {\cal A}^{\pm}) \cr
&& \tilde {\mathbb V}^{\pm (3)}(\mu) = \mu^2 e_{NN} + {\mu \over 2} ({\cal A}^{\pm}- \bar { \cal A}^{\pm}) + {1\over 2} ({\cal B}^{\pm} + \bar {\cal B}^{\pm})
\ea
(we have divided all the results coming from (\ref{tvv}) with an overall ${1\over 2}$ factor); we also define:
\ba
&& {\cal A}^- = \sum_{j\neq N} (\kappa {\mathbb P}_{Nj} -W_{Nj}^{+(1)})e_{Nj} +\sum_{j\neq N} W_{jN}^{-(1)} e_{jN} \cr
&& \bar {\cal A}^- = \sum_{j\neq N} (\kappa {\mathbb P}_{jN} -W_{jN}^{+(1)})e_{jN} + \sum_{j\neq N} W_{Nj}^{-(1)} e_{Nj}\cr
&& {\cal A}^+ = \sum_{j\neq N} (\kappa {\mathbb P}_{jN} +W_{jN}^{-(1)})e_{jN} -\sum_{j\neq N} W_{Nj}^{+(1)} e_{Nj} \cr
&& \bar {\cal A}^+ = \sum_{j\neq N} (\kappa {\mathbb P}_{Nj} + W_{Nj}^{-(1)})e_{Nj} - \sum_{j\neq N} W_{jN}^{+(1)} e_{jN}.
\ea
\ba
{\cal B}^{-} &=& \sum_{j\neq N} \Big (i W_{Nj}^{+(1)'}- \kappa \sum_{i \neq N}W_{Ni}^{+(1)} {\mathbb P}_{ij} - \kappa^2 {\mathbb P}_{NN} {\mathbb P}_{Nj} + \kappa {\mathbb P}_{NN} W_{Nj}^{+(1)}  \Big ) e_{Nj} + i \sum_{j\neq N} W_{jN}^{-(1)'}e_{jN} \cr
&+&   \sum_{i, j  \neq N} \Big ( \kappa W_{jN}^{-(1)} {\mathbb P}_{Ni}  - W_{jN}^{-(1)} W_{Ni}^{+(1)} \Big )e_{ji}-
\sum_{ j  \neq N} \Big ( \kappa {\mathbb P}_{Nj} W_{jN}^{-(1)} - W_{Nj}^{+(1)} W_{jN}^{-(1)}\Big )e_{NN} \cr
\bar {\cal B}^{-} &=& \sum_{j\neq N} \Big (  i W_{jN}^{+(1)'}- \kappa \sum_{l \neq N} {\mathbb P}_{jl}  W_{lN}^{+(1)}  + \kappa^2 {\mathbb P}_{jl} {\mathbb P}_{lN}  + \kappa  W_{jN}^{+(1)} {\mathbb P}_{NN} \Big ) e_{jN} + i \sum_{j\neq N} W_{Nj}^{-(1)'}e_{Nj} \cr
&+&   \sum_{i, j  \neq N} \Big ( \kappa {\mathbb P}_{jN} W_{Ni}^{-(1)} - W_{jN}^{+(1)} W_{Ni}^{-(1)}\Big )e_{ji}+
\sum_{ j  \neq N} \Big ( -\kappa W_{Nj}^{-(1)} {\mathbb P}_{jN}  +  W_{Nj}^{-(1)} W_{jN}^{+(1)} \Big )e_{NN}.
\nonumber \\
\ea
${\mathbb P}_{ij}$ are associated to defect and are defined in (\ref{ll1}), (\ref{ll2}).

Similarly, we define ${\cal B}^+, \bar {\cal B}^+$ as:
\ba
{\cal B}^{+} &=& \sum_{j\neq N} \Big (i W_{jN}^{-(1)'}- \kappa \sum_{l \neq N} {\mathbb P}_{jl}  W_{lN}^{-(1)}  - \kappa^2 {\mathbb P}_{jN} {\mathbb P}_{NN}  - \kappa W_{jN}^{-(1)} {\mathbb P}_{NN}   \Big ) e_{jN} + i \sum_{j\neq N} W_{Nj}^{+(1)'}e_{Nj} \cr
&-& \sum_{i, j  \neq N} \Big ( \kappa {\mathbb P}_{jN} W_{Ni}^{+(1)} + W_{jN}^{-(1)} W_{Ni}^{+(1)}\Big )e_{ji}+
\sum_{ j  \neq N} \Big ( \kappa W_{Nj}^{+(1)} {\mathbb P}_{jN} + W_{Nj}^{+(1)}W_{jN}^{-(1)} \Big )e_{NN}  \cr
\bar {\cal B}^{+} &=& \sum_{j\neq N} \Big (i W_{Nj}^{-(1)'}+ \kappa \sum_{l \neq N}W_{Nl}^{-(1)} {\mathbb P}_{lj} + \kappa^2 \sum_{l\neq j} {\mathbb P}_{Nl} {\mathbb P}_{lj} + \kappa {\mathbb P}_{NN} W_{Nj}^{+(1)}  \Big ) e_{Nj} + i \sum_{j\neq N} W_{jN}^{+(1)'}e_{jN} \cr
&-& \sum_{i, j  \neq N} \Big ( \kappa W_{jN}^{+(1)} {\mathbb P}_{Ni}  + W_{jN}^{+(1)} W_{Ni}^{-(1)} \Big )e_{ji}+
\sum_{ j  \neq N} \Big ( \kappa {\mathbb P}_{Nj} W_{jN}^{+(1)} + W_{Nj}^{-(1)} W_{jN}^{+(1)}\Big )e_{NN}. \nonumber \\
\ea
Due to continuity requirements around the defect point (\ref{cont})
we obtain the following sewing conditions:
\ba
&& \kappa {\mathbb P}_{Nj} = W_{Nj}^{+(1)} - W_{Nj}^{-(1)} \cr
&& \kappa {\mathbb P}_{jN} = W_{jN}^{+(1)} - W_{jN}^{-(1)}. \label{glue}
\ea
The conditions associated to the 3d order are given by
\ba
&& W_{Nj}^{+(1)'} -  W_{Nj}^{-(1)'} = i\kappa \Big( W_{Nj}^{+(1)} {\mathbb P}_{NN} - \sum_{l\neq N} {\mathbb P}_{lj} W_{Nl}^{+(1)} - \kappa {\mathbb P}_{NN} {\mathbb P}_{Nj}\Big ) \label{1} \\
&& W_{jN}^{+(1)'} -  W_{jN}^{-(1)'} = i\kappa \Big(W_{jN}^{+(1)} {\mathbb P}_{NN}  - \sum_{l\neq N} {\mathbb P}_{jl} W_{lN}^{+(1)} + \kappa  \sum_{l\neq N}{\mathbb P}_{jl} {\mathbb P}_{lN}\Big ), \label{1b}
\ea
and
\ba
&& W_{jN}^{+(1)'} -  W_{jN}^{-(1)'} = i\kappa \Big({\mathbb P}_{NN} W_{jN}^{-(1)} - \sum_{l \neq N} W_{lN}^{-(1)} {\mathbb P}_{jl} + \kappa  {\mathbb P}_{jN} {\mathbb P}_{NN}\Big ) \label{2}\\
&& W_{Nj}^{+(1)'} -  W_{Nj}^{-(1)'} = i\kappa \Big({\mathbb P}_{NN} W_{Nj}^{+(1)} - \sum_{l\neq N} W_{Nl}^{+(1)} {\mathbb P}_{lj} - \kappa {\mathbb P}_{NN} {\mathbb P}_{Nj}\Big ). \label{2b}
\ea
As expected the odd quantities are non-trivial, associated to the corresponding conserved quantities
i.e. number of particles, Hamiltonian etc, whereas the even ones are naturally zero. Notice than (\ref{1b}), (\ref{2}) and (\ref{1}), (\ref{2b}) are compatible as one would expect due to the gluing conditions (\ref{glue}). Moreover, the sewing conditions derived above are compatible with the Hamiltonian action in accordance to the relevant generic proposition proved in \cite{avan-doikou1} for systems associated to the Yangian classical $r$-matrix.

The next natural step is to identify the equations of motion for the system. It is clear that in the bulk the equations of motion associated to the Hamiltonian are given by the familiar expressions for the vector NLS model:
\be
i{\partial \psi^{\pm}_{i}(x,t) \over
\partial t} = - {\partial^{2} \psi^{\pm}_{i}(x,t) \over
\partial^2 x}+2\kappa \sum_{j=1}^{N-1}|\psi^{\pm}_{j}(x,t)|^2 \psi^{\pm}_{i}(x,t),
~~~~i \in \{1, \ldots, N-1 \}, ~~~x \neq x_0. \label{nls}
\ee

Recall that the Hamiltonian of the system was explicitly derived in \cite{doikou-vectornls} as:
\ba
&& {\cal I}_3= -2i\kappa \int_{0}^{x_0} dx\ \sum_{i=1}^{N-1} \Big (\kappa
|\psi^-_i(x)|^2 \sum_k|\psi^-_k(x)|^2  +\psi_i^{-'}(x) \bar \psi_i^{-'}(x)
\Big ) + 2i \kappa \sum_i \psi_i^-(x_0) \bar \psi_i^{-'}(x_0) \cr
&& -2i\kappa \int_{x_0}^{L} dx\ \sum_{i=1}^{N-1} \Big (\kappa
|\psi^+_i(x)|^2 \sum_k|\psi^+_k(x)|^2  +\psi_i^{+'}(x) \bar \psi_i^{+'}(x)
\Big )-  2 i \kappa \sum_i \psi_i^+(x_0) \bar \psi_i^{+'}(x_0)
\cr && -2i\kappa \sum_{j} (\psi_j(0)\bar \psi(0))' + {\cal D}_3 + \bar {\cal D}_3  +
\kappa^2 \sum_{j\neq N, i}{\mathbb P}_{Ni} {\mathbb P}_{ij} W_{jN}^{+(1)} + \kappa^2 \sum_{i\neq N, j} W_{Ni}^{-(1)}{\mathbb P}_{ij} {\mathbb P}_{jN} + \kappa^3 \sum_{i \neq N, j} {\mathbb P}_{Ni} {\mathbb P}_{ij} {\mathbb P}_{jN} \cr
&& \label{first2}
\ea
where ${\cal D}_3$ is defined as follows
\be
{\cal D}_3 = D_3 -D_1 D_2 +{D_1^3 \over 3}
\ee
\ba
&& D_1 = \kappa {\mathbb P}_{NN} \cr
&& D_2 = \sum_{j\neq N} W_{Nj}^+ W_{jN}^{+} - \kappa \sum_{j\neq N} W_{Nj}^{+(1)}{\mathbb P}_{jN} + \kappa \sum_{j\neq N} {\mathbb P}_{Nj} W_{jN}^{-(1)} - \sum_{j\neq N} W_{Nj}^{+(1)}W_{jN}^{-(1)} \cr
&& D_3 = \sum_{j \neq N} W_{Nj}^{+(1)}W_{jN}^{+(2)}+\sum_{j \neq N} W_{Nj}^{+(2)}W_{jN}^{+(1)} - \sum_{i\neq N  i, j \neq N} W_{Ni}^{+(1)} W_{ij}^{+(1)}W_{jN}^{+(1)}  - \sum_{j \neq N} W_{Nj}^{+(1)}W_{jN}^{-(2)} \cr
&&+ \kappa \sum_{j \neq N} {\mathbb P}_{Nj} W_{jN}^{-(2)}
+ \sum_{i\neq N, j \neq N} W_{Ni}^{+(1)} W_{ij}^{+(1)}W_{jN}^{-(1)} - \sum_{j\neq N} W_{Nj}^{+(2)}W_{jN}^{-(1)} -\sum_{i\neq N,j\neq N} W_{Nj}^{+(1)}{\mathbb P}_{ij} W_{jN}^{-(1)}. \label{dd} \cr
&& -\kappa \sum_{j\neq N}W_{Nj}^{+(2)}{\mathbb P}_{jN} + \kappa \sum_{i\neq N, j} W_{Ni}^{+(1)} W_{ij}^{+(1)} {\mathbb P}_{jN}
\ea
and
\be
\bar {\cal D}_3(W_{ij}^{\pm (n)}) = {\cal D}_3((-1)^n W_{ij}^{\mp (n)}). \label{refl}
\ee

However, special care is taken on the defect point (\ref{zcd}), then the following time evolution for the defect degrees of freedom is obtained:
\be
\dot {\mathbb P}_{ij} = {1\over 2}({\cal B}^+_{il} + \bar {\cal B}^+_{il}){\mathbb P}_{lj} -{1\over 2}{\mathbb P}_{il}({\cal B}^-_{lj} + \bar {\cal B}^-_{lj}).
\ee
It is worth noting that in the periodic case studied in \cite{doikou-vectornls} the sewing condtions were used in order to eliminate the $\lambda$ dependent terms in the zero curvature condition on the defect point. In the present case however these terms disappear automatically due to the various contributing terms emerging from the ``double'' monodromy matrix ${\cal T}$.
Comparing also the equations of motion above with the ones extracted in the periodic case \cite{doikou-vectornls} we conclude that are somehow ``double'', due to the presence of the reflecting boundary conditions.

\subsection{The twisted Yangian}

Let us now briefly describe in this subsection  the SNP case. This case is not particularly ``inspiring'' in the sense that it does not provide any ``unexpected'' bulk behavior due to the presence of the integrable boundaries. Nevertheless, it is worth studying it for completeness as well as for showing that the presence of boundaries does not always leads to drastic effects in the bulk. Moreover, this study further confirms the results presented in \cite{doikou-vectornls}.

It is technically  more tractable to consider the discrete vector NLS and then take the suitable continuum limit. Admittedly, working directly on the continuum case it is much more complicated mainly due to various contributions from the diagonal $Z$ terms.
The discrete monodromy matrix in the presence of a single defect on the $n$-th cite:
\be
T_0(L, 1; \lambda) = {\mathbb L}_{0L}(\lambda)\ldots \tilde {\mathbb L}_{0n}(\lambda) \ldots {\mathbb L}_{01}(\lambda)
\ee
In general, we introduce the notation:
\be
T_0(i, j;\lambda) = {\mathbb L}_{0i} \ldots {\mathbb L}_{0j}(\lambda) ~~~~i > j.
\ee
it is clear that in the case $i > n > j$ the $\tilde {\mathbb L}$ matrix acts on the $n^{th}$ site of the one dimensional discrete system.

Both ${\mathbb L}$ and $\tilde {\mathbb L}$, and consequently $T$ satisfy the classical quadratic algebra (\ref{rtt2}).
Where the ${\mathbb L}$ matrix of the discrete vector NLS model is given as
\be
{\mathbb L}(\lambda) = (i \lambda + \kappa {\mathbb N} )\ e_{NN} + \sum_{l = 1}^{N-1}e_{ll} + \sum_{l=1}^{N-1}(\phi_{l}\ e_{lN} + \psi_{l}\ e_{Nl}).
\ee
Due to the quadratic algebra we obtain:
\ba
&& \Big \{\phi_{k},\ \psi_{l} \Big \} = i \delta_{kl}\cr
&& \Big \{ {\mathbb N},\ \phi_k \Big  \} = -i \phi_{k} \cr
&& \Big \{ {\mathbb N},\ \psi_k  \Big \} = i \psi_{k}.
\ea

The defect Lax operator is
\ba
\tilde {\mathbb L}_n(\lambda) = \lambda + \kappa {\mathbb P}^{(n)}.
\ea
the entries of the ${\mathbb P}$ matrix ${\mathbb P}_{ij}$ satisfy the classical $\mathfrak{gl}_N$ algebra:
\be
\Big \{ {\mathbb P}_{ij}^{(n)},\ {\mathbb P}_{kl}^{(n)} \Big \}
= {\mathbb P}_{il}^{(n)} \delta_{jk}- {\mathbb P}_{jk}^{(n)} \delta_{il}.
\ee

As already mentioned we focus  on the SNP case that is we consider the representation of the classical twisted Yangian (\ref{ty}). Expansion of $tr {\cal T}(\lambda)$, --${\cal T}$ defined in (\ref{rep2}), but now $T$ is the discrete monodromy matrix-- will lead to the local integrals of motion (see also \cite{doikou-fioravanti-ravanini} for relevant results)
\ba
{\cal I}^{(2)}_d &=& \kappa \sum_{l=1}^{N-1} \sum_{j \neq n, n-1} \psi_l^{(j+1)} \phi_l^{(j)} + i \kappa \sqrt{\kappa} \sum_{l=1}^{N-1}\Big ( \psi_l^{(n+1)} {\mathbb P}_{lN}^{(n)} + {\mathbb P}_{Nl}^{(n)} \phi^{(n-1)}_l \Big ) - \kappa^2 \sum_{j=1}^{N-1}({\mathbb N}^{(j)})^2 \cr
&+& \kappa \sum_{l=1}^{N-1} \psi_l^{(n+1)} \phi_l^{(n-1)} - \kappa \sum_{l=1}^{N-1}\Big (\phi_l^{(L)}\phi_l^{(L)} + \psi_l^{(1)}\psi_l^{(1)} \Big ). \nonumber\\
\ea
By taking the continuum limit as (see also \cite{avan-doikou1} and references therein)
\ba
&& \Big ( \psi_l^{(j)},\ \phi^{(j)}_l \Big ) \to \Big ( \psi^+_l(x),\ \bar \psi^+_l(x) \Big ), ~~~~~~j > n, ~~~~x > x_0\cr
&& \Big ( \psi_l^{(j)},\ \phi^{(j)}_l \Big ) \to \Big ( \psi^-_l(x),\ \bar \psi^-_l(x) \Big ), ~~~~~~j < n, ~~~~x < x_0 \label{climit}
\ea
we obtain the continuum vector NLS momentum \cite{doikou-vectornls}
\ba
   && {\cal I}_2 =- \kappa \int_{0}^{x_0} dx\ \sum_{i=1}^{N-1} \Big (\bar \psi^-_i(x)
\psi^{-'}_i(x)- \psi^-_i(x) \bar \psi^{-'}_i(x) \Big ) +\kappa \sum_i\psi_i^-(x_0)\bar \psi_i^-(x_0)
\cr
&& -\kappa
 \int_{x_0}^{L} dx\ \sum_{i=1}^{N-1} \Big (\bar \psi^+_i(x)
\psi^{+'}_i(x)- \psi_i^+(x) \bar \psi^{'+}_i(x) \Big ) - \kappa \sum_i \psi_i^+(x_0)\bar \psi_i^+(x_0)  \cr && +\kappa \sum_i\psi_i^-(0)\bar \psi_i^-(0) +  \kappa \sum_i \psi_i^-(0) \psi_i^-(0) +2 {\cal D}_2  \cr &&
\ea
${\cal D}_2$ is defined as ${\cal D}_2 = D_2 - {D_1^2 \over 2}$,
and $D_1,\ D_2$ are given in (\ref{dd}).
The charge ${\cal I}_2$ provides indeed the momentum of the system:
\be
{\cal P} = - { {\cal I}_2 \over 2 i \kappa }.
\ee
In general, as discussed in \cite{doikou-vectornls} in this case only the even charges survive.

Let us now formulate the corresponding discrete Lax pairs.
In the discrete case the zero curvature condition is expressed as (see also \cite{ft}):
\ba
&& {d {\mathbb L}_{j}(\lambda)\over dt} = {\mathbb A}_{j+1}(\lambda, \mu)\ {\mathbb L}_j(\lambda)- {\mathbb L}_j(\lambda)\ {\mathbb A}_{j}(\lambda, \mu),
~~~~~j \neq n, \cr
&& {d \tilde {\mathbb L}_{n}(\lambda)\over dt} = {\mathbb A}_{n+1}(\lambda, \mu)\ \tilde {\mathbb L}_n(\lambda) - \tilde {\mathbb L}_n(\lambda)\ {\mathbb A}_{n}(\lambda,\mu).\label{zero-bulk}
\ea
As in the continuum case the time components of the Lax pair can be identified:
\be
{\mathbb A}_{j}(\lambda, \mu) = t^{-1}(\lambda)\ tr_a \Big ({\cal M}_{ab}^{(j)}(\lambda, \mu) +
{\cal M}_{ab}^{*(j)}(\lambda, \mu) \Big)
\ee
and
\be
A_j(\lambda, \mu) = \sum_m {A_j^{(m)}(\mu) \over \lambda^m}.
\ee
We also define the discrete quantities (we have considered for simplicity $K= \bar K \propto {\mathbb I}$):
\be
{\cal M}_{ab}^{(j)}(\lambda, \mu)= T_a(N, j;\lambda)\  r_{ab}(\lambda-\mu)\ T_a(j-1, 1;\lambda)\ T_a^{t_a}(N, 1; -\lambda), ~~~j \neq n,\ n+1.
\ee
Around the defect point we define:
\ba
&& {\cal M}_{ab}^{(n)}(\lambda, \mu) = T_a(L, n+1;\lambda)\ \tilde
{\mathbb L}_{a n}(\lambda)\ r_{ab}(\lambda-\mu)\ T_a(n-1, 1;\lambda)\ T_a^{t_a}(N,1; -\lambda) \cr
&& {\cal M}_{ab}^{(n+1)}(\lambda, \mu) = T_a(L, n+1;\lambda)\ r_{ab}(\lambda-\mu)\ \tilde
{\mathbb L}_{a n}(\lambda)\ T_a(n-1, 1;\lambda)\ T_a^{t_a}(N,1; -\lambda)
\ea
\be
{\cal M}_{ab}^{*}(\lambda, \mu
) = {\cal M}_{ab}^{t_a}(-\lambda, \mu).
\ee

Expansion in powers of ${1 \over \lambda}$ of ${\mathbb A}_{j} = \sum_m {{\mathbb A}_{j}^{(m)} \over \lambda^m}$ will provide the time component of the Lax pairs associated to each integral of motion. The first non trivial is the one associated to the momentum:
\be
{\mathbb A}_{j}^{(2)}(\mu) = i\mu\ e_{NN} + \sqrt{\kappa} \sum_{l\neq N} \Big ( \psi_l^{(j)}\ e_{Nl} +  \phi_l^{(j)}\ e_{lN}\Big ) , ~~~~j\neq n,\ n+1
\ee
and around the defect point we derive:
\ba
&& {\mathbb A}_{n}^{(2)}(\mu) =  i\mu \ e_{NN} + \sqrt{\kappa}\sum_{l\neq N}  \Big (\psi_l^{(n+1)} +
i \kappa {\mathbb P}_{Nl}^{(n)}\Big)\ e_{Nl} +  \sqrt{\kappa} \sum_{l\neq N} \phi_l^{(n-1)}e_{lN} \cr
&& {\mathbb A}_{n+1}^{(2)}(\mu)= i \mu\ e_{NN} + \sqrt{\kappa}\sum_{l\neq N} \psi_l^{(n+1)}  e_{Nl} +  \sqrt{\kappa} \sum_{l\neq N} \Big ( \phi_l^{(n-1)}+
i \kappa {\mathbb P}_{lN}^{(n)}\Big)\ e_{lN}.
\ea

We are now taking the continuum limit exploiting (\ref{climit}) and
\ba
&& {\mathbb A}_j^{(2)} \to {\mathbb V}^{+(2)}(x), ~~~~~~j > n+1, ~~~~x > x_0 \cr
&& {\mathbb A}_j^{(2)} \to {\mathbb V}^{-(2)}(x), ~~~~~~j < n, ~~~~x < x_0 \cr
&& {\mathbb A}_{n+1}^{(2)} \to \tilde {\mathbb V}^{+(2)}(x_0),\cr
&& {\mathbb A}_{n}^{(2)} \to \tilde {\mathbb V}^{-(2)}(x_0),
\ea
we end up with the following expressions on the continuum time components of the Lax pair associated to the momentum:
\be
{\mathbb V}^{\pm(2)}(\mu; x) = i\mu\ e_{NN}+ \sqrt{\kappa} \sum_{l \neq N} \Big (\psi^{\pm}_l(x)\ e_{Nl}
+ \bar \psi^{\pm}_l(x)\ e_{lN} \Big )
\ee
and on the defect point:
\ba
&& \tilde {\mathbb V}^{-(2)}(\mu; x_0) = i\mu\ e_{NN}+ \sqrt{\kappa} \sum_{l \neq N}
\Big (\psi^+_l(x_0) + i\kappa {\mathbb P}_{Nl}(x_0)  \Big )\ e_{Nl}
+ \bar \psi^{-}_l(x_0)\ e_{lN} \cr
&& \tilde {\mathbb V}^{+(2)}(\mu; x_0) = i \mu\ e_{NN}+ \sqrt{\kappa} \sum_{l \neq N} \psi^{+}_l(x_0)e_{Nl}
+ \Big ( \bar \psi^{-}_l(x_0) + i \kappa {\mathbb P}_{lN}(x_0) \Big )\ e_{lN}.
\ea
Then continuity conditions imposed around the defect point (\ref{cont}) lead to the sewing conditions extracted in the periodic case \cite{doikou-vectornls} i.e.
\ba
&& \sqrt{\kappa}\ {\mathbb P}_{Nl} = i (\psi^+_l -  \psi_l^-) \cr
&& \sqrt{\kappa}\ {\mathbb P}_{Nl} =- i (\bar \psi^+_l -  \bar \psi_l^-).
\ea
That is the presence of non trivial integrable boundary conditions
in the SP case does not really affect the behavior of the defect as expected.
With this we conclude the derivation of Lax pairs in the simultaneous
presence of boundaries and defects for vector NLS model.

\section{Discussion}

The Lax pairs in the simultaneous presence of point like defects and integrable boundaries were constructed. Based on this construction the vector NLS model was investigated. We have used the vector NLS model as a prototype in order to exemplify our results as well as to make contact with recent relevant results found in \cite{doikou-vectornls}. The vector NLS model displays a highly non-trivial behavior compared to the single NLS due to the fact that it is ruled by the high rank $\mathfrak{gl_N}$ algebra, and this allows the existence of two distinct types of boundary conditions associated to the reflection algebra and the twisted Yangian as opposed to the self-dual $\mathfrak{sl}_2$ case, where basically the two algebras coincide. It is worth noting that we have basically considered $K \propto {\mathbb I}$ for simplicity but without loss of generality. Another choice of $K$ matrices would have no effect whatsoever on the bulk Lax pair or the Lax pairs associated to the defect point --and this is one of the key points in the present derivation--, but it would only add some extra boundary terms in the relevant local integrals of motion (see also \cite{doikou-vectornls}). Thus what is important in the present description is in fact the choice of the underlying classical algebra whether this is the reflection algebra or the twisted Yangian.

Although this construction is the first significant step towards understanding the behavior of such systems there are still various points in this context to further explore. For instance the solution of the relevant equations of motion exploiting the integrable sewing conditions across the defect point is a pertinent issue to address. It would be also very interesting to identify the classical scattering data of the solitonic solutions with the defect derived via the inverse scattering method and compare with the corresponding quantum results derived in numerous works (see e.g. \cite{konle, BCZ3, fus, doikou-karaiskos, doikou-defects}). The relevant solitonic solutions in this context can be extracted via the associated B\"{a}cklund transformations, which in turn can be derived through the solution of the inverse scattering \cite{wadati}; this is a particulary interesting issue especially with regard to the defect point, and its connection with the integrable sewing conditions. Another significant question worth exploring would be the identification of the explicit form of the Jost functions in the presence of defects and the formulation of the associated Gelfand-Levitan-Marchenko equations at both classical and quantum level \cite{ft,sogo-wadati}.

This line of attack can be implemented into various classical systems and connection with relevant results from the quantum point of view identified via the Bethe ansatz formulation (see e.g. \cite{doikou-karaiskos, doikou-defects}) can be made. Finally, a similar description in the context of $A_n^{(1)}$ affine Toda field theories would also yield highly non-trivial findings on the behavior of the system. The boundary affine Toda field theories associated to the twisted Yangian and the reflection algebras was studied in \cite{doikou-toda} and highly non trivial boundary behavior was exhibited in the case of reflecting boundaries. Similarly, intricate phenomena associated to the defects are expected in the case of the SP boundary conditions. Hopefully, all the aforementioned issues will be addressed in future investigations.


\begin{thebibliography}{99}


\bibitem{delmusi}
G. Delfino, G. Mussardo and P. Simonetti, Phys. Lett. {\bf B328} (1994) 123, {\tt hep-th/9403049}.

\bibitem{delmusi1}
G. Delfino, G. Mussardo and P. Simonetti, Nucl. Phys. {\bf B432} (1994) 518, {\tt hep-th/9409076}.

\bibitem{konle}
R. Konik and A. LeClair, Nucl. Phys.  {\bf B538} (1999) 587; {\tt hep-th/9793985}.

\bibitem{BCZ1}
P. Bowcock, E. Corrigan and C. Zambon, JHEP 01 (2004) 056, {\tt hep-th/0401020}

\bibitem{BCZ2}
P. Bowcock, E. Corrigan and C. Zambon, JHEP 08 (2005) 023, {\tt hep-th/0506169}.

\bibitem{BZ} E. Corrigan, C. Zambon, Nonlinearity 19 (2006) 1447, {\tt nlin/0512038}.

\bibitem{BCZ3}
E. Corrigan and C. Zambon, JHEP 07 (2007) 001, {\tt arXiv:0705.1066 [hep-th]}.

\bibitem{BCZ4}
E. Corrigan and C. Zambon, J. Phys. {\bf A42} (2009) 475203, {\tt arXiv:0908.3126 [hep-th]}.

\bibitem{cozanls}
E. Corrigan and C. Zambon, Nonlinearity {\bf 19} (2006) 1447, {\tt nlin/0512038}.

\bibitem{fus}
E. Corrigan and C. Zambon, J. Phys. {\bf A43} (2010) 345201, {\tt arXiv:1006.0939 [hep-th]}.

\bibitem{haku}
I. Habibullin and A. Kundu, Nucl. Phys. {\bf B795} (2008) 549, {\tt arXiv:0709.4611 [hep-th]}.

\bibitem{caudr}
V. Caudrelier, IJGMMP vol.5, {\bf No. 7} (2008) 1085, {\tt arXiv:0704.2326 [math-ph]}.

\bibitem{nemes}
F. Nemes, Int. J. Mod. Phys. {\bf A25} (2010) 4493; {\tt arXiv:0909.3268 [hep-th]}.

\bibitem{weston}
R. Weston, {\tt arXiv:1006.1555 [math-ph]}.

\bibitem{annecydef1}
M. Mintchev, E. Ragoucy and P. Sorba, Phys. Lett. {\bf B547} (2002) 313, {\tt hep-th/0209052};\\
M. Mintchev, E. Ragoucy and P. Sorba, J. Phys. {\bf A36} (2003) 10407, {\tt hep-th/0303187}.

\bibitem{annecydef2}
V. Caudrelier, M. Mintchev and E. Ragoucy, J. Phys. {\bf A37} (2004) L367, {\tt hep-th/0404144}.

\bibitem{agui}
A.R. Aguirre, T.R. Araujo, J.F. Gomes and A.H. Zimerman, JHEP 12 (2011) 56, {\tt arXiv:1110.1589 [hep-th]}.

\bibitem{agui2}
A.R. Aguirre, J. Phys. {\bf A 45} (2012) 205205, {\tt arXiv:1111.5249 [math-ph]}.

\bibitem{avan-doikou1}
A. Doikou, Nucl. Phys. {\bf B854} (2012) 153, {\tt arXiv:1106.1602, [hep-th]};\\
J. Avan and A. Doikou, JHEP 01 (2012) 040, {\tt arXiv:1110.4728 [hep-th]}.

\bibitem{avan-doikou}
J. Avan and A. Doikou, JHEP 11 (2012) 008,  {\tt arXiv:1205.1661 [hep-th]}.

\bibitem{doikou-karaiskos-sigma}
A. Doikou and N. Karaiskos, Nucl. Phys. {\bf B867} (2013) 872, {\tt  arXiv:1207.5503 [hep-th]}.

\bibitem{doikou-karaiskos}
A. Doikou and N. Karaiskos, JHEP 02 (2013) 142, {\tt arXiv:1212.0195 [math-ph]}.

\bibitem{doikou-defects}
A. Doikou, JHEP 08 (2013) 103, {\tt  arXiv:1304.5901 [hep-th]};\\
A. Doikou, Nucl. Phys. {\bf B877} (2013) 885, {\tt arXiv:1307.2752 [math-ph]};\\
A. Doikou, {\tt arXiv:1308.1790 [math-ph]}.

\bibitem{doikou-vectornls}
A. Doikou, Nucl. Phys. {\bf B884} (2014) 142,  {\tt arXiv:1312.4786}.

\bibitem{ft}
L.D. Faddeev and L.A. Takhtakajan, {\it Hamiltonian Methods in the Theory of Solitons},
(1987) Springer-Verlag.

\bibitem{sts}
M.A. Semenov-Tjan-Shanskii, Funct. Anal. Appl. {\bf 17} (1983), 259.

\bibitem{sklyanin}
E.K. Sklyanin, Funct. Anal. Appl. {\bf 21} (1987) 164;\\
E.K. Sklyanin, J. Phys. {\bf A21} (1988) 2375.

\bibitem{olshanski} 
G.I. Olshanski, {\it Twisted Yangians and inﬁnite-dimensional classical Lie algebras in
‘Quantum Groups’} (P.P. Kulish, Ed.), Lecture notes in Math. 1510, Springer (1992) 103.

\bibitem{doikou-twisted}
A. Doikou, J. Phys. {\bf A33} (2000) 8797, {\tt hep-th/0006197}.

\bibitem{avan-doikou-lax}
J. Avan and A. Doikou, Nucl. Phys. {\bf B800} (2008) 591, {\tt arXiv:0710.1538 [hep-th]};\\
J. Avan and A. Doikou, Nucl. Phys. {\bf B821} (2009) 481, {\tt arXiv:0809.2734 [hep-th]}.

\bibitem{sklyanin2} 
E. Sklyanin, J. Sov. Math. 19 (1982) 1564.

\bibitem{yang}
C.N. Yang, Phys. Rev. Lett. {\bf 19} (1967) 1312.

\bibitem{doikou-fioravanti-ravanini}
A. Doikou, D. Fioravanti and F. Ravanini,  Nucl. Phys. {\bf B790} (2008) 465, {\tt  arXiv:0706.1515 [hep-th]}.

\bibitem{doikou-toda}
A. Doikou, JHEP 0805 (2008) 091, {\tt  arXiv:0803.0943 [hep-th]};\\
J. Avan and A. Doikou, Nucl. Phys. {\bf B821} (2009) 481, {\tt  arXiv:0809.2734 [hep-th]}.

\bibitem{wadati}
K. Konno and M. Wadati, Progr. Theor. Phys. {\bf 53} (1976) 1652.

\bibitem{sogo-wadati}
K. Sogo and M. Wadati, Progr. Theor. Phys. {\bf 69} (1983) 431.




\end{thebibliography}
\end{document}